# Myocardial T$_1$ mapping at 5T using multi-inversion recovery real-time spoiled GRE

Linqi Ge[1,2], Yihang Zhang[1,3], Huibin Zhu[1,4], Lang Zhang[1,3], Yihang Zhou[1], Haifeng Wang[1], Dong Liang[1], Hairong Zheng[1], Yanjie Zhu[1]

[1]Shenzhen Institutes of Advanced Technology, Chinese Academy of Sciences, 1068 Xueyuan Avenue, Guangdong, Shenzhen, China

[2]University of Chinese Academy of Sciences, 80 Zhongguancun East Road, Beijing, China

[3]College of Pharmacy and Bioengineering, Chongqing University of Technology, Chongqing, China

[4]School of Mathematical Science, Inner Mongolia University, Hohhot, China

**\*Correspondence to:**

Yanjie Zhu, Ph.D.

Paul C. Lauterbur Research Centre for Biomedical Imaging

Shenzhen Institutes of Advanced Technology

Chinese Academy of Sciences, Shenzhen, Guangdong, China, 518055

Tel: (86) 755-86392243

Fax: (86) 755-86392299

Email: yj.zhu@siat.ac.cn

**ABSTRACT**

**Objective:** To develop an accurate myocardial $T_1$ mapping technique at 5T using Look-Locker-based multiple inversion-recovery with the real-time spoiled gradient echo (GRE) acquisition.

**Approach:** The proposed $T_1$ mapping technique (mIR-rt) samples the recovery of inverted magnetization using the real-time GRE and the images captured during diastole are selected for $T_1$ fitting. Multiple-inversion recoveries are employed to increase the sample size for accurate fitting. The $T_1$ mapping method was validated using Bloch simulation, phantom studies, and in 16 healthy volunteers at 5T.

**Main results:** In both simulation and phantom studies, the $T_1$ values measured by mIR-rt closely approximate the reference $T_1$ values, with errors less than 3%, while the conventional MOLLI sequence underestimates $T_1$ values. The myocardial $T_1$ values at 5T are 1553 ± 52 ms, 1531 ± 53 ms, and 1526 ± 60 ms (mean ± standard deviation) at the apex, middle, and base, respectively. The $T_1$ values measured by MOLLI (1350 ± 48 ms, 1349 ± 47 ms, and 1354 ± 45 ms at the apex, middle, and base) were significantly lower than those of mIR-rt with $p < 0.001$ for all three layers. The mIR-rt sequence method used in our study provides high reproducibility, particularly in the middle slices, supporting its practical relevance for myocardial $T_1$ mapping.

**Significance:** The proposed method is feasible for myocardial $T_1$ mapping at 5T and provides better accuracy than the conventional MOLLI sequence.

**Keywords:** Myocardial $T_1$ mapping, 5T, Look-Locker, mIR-rt

# 1. Introduction

Myocardial $T_1$ mapping can depict subtle variations in myocardium, allowing the detection of myocardial amyloidosis, iron overload, and myocardial infarction without using contrast agents [1, 2]. It has been widely used in clinical diagnosis and research at 1.5T and 3T. In recent years, although ultra-high field MR scanners, including 5T and 7T, have become popular around the world, cardiac imaging at ultra-high fields is still in its infancy.

Various techniques have been employed to quantify myocardial $T_1$ relaxation time, each having its own merits and limitations [3]. The most commonly used technique is the Modified Look-Locker Inversion Recovery (MOLLI) sequence and its derivations [4-6]. In MOLLI, single images are intermittently acquired during diastole gated by electrocardiogram in 3 to 5 heartbeats following the inversion recovery (IR) pulse, resulting in images spaced along $T_1$ recovery curves at RR interval. Then the images acquired after multiple IRs are concatenated together for $T_1$ curve fitting using the Look-Locker model. However, MOLLI-based sequences would underestimate $T_1$ values at 5T. This underestimation is attributed to the longer T1 relaxation times at higher field strengths, which require more time for magnetization recovery between inversion pulses. Since MOLLI sequences assume complete recovery of magnetization within the acquisition window, the insufficient recovery time at 5T leads to inaccuracies in T1 estimation. Studies[7] have shown that at ultra-high field strengths, such as 7T, myocardial T1 values are significantly longer, necessitating adjustments to the acquisition protocol for accurate measurement. Saturation Recovery Single-Shot Acquisition (SASHA) [8] provides an alternative to overcome this issue but suffers from limited signal-to-noise ratio (SNR) and precision. The longer $T_1$ at 5T makes this issue even worse. STONE sequence [9] can achieve accurate $T_1$ measurement with high SNR but must acquire five slices together.

Inversion recovery with real-time sampling has been a cornerstone of $T_1$ mapping for decades[10]. Traditional Cartesian sampling remains the most widely used approach due to its simplicity and compatibility with standard reconstruction algorithms. However, recent advancements in radial acquisition and model-based reconstruction offer promising alternatives[11, 12]. Radial trajectories, with their inherent motion robustness, have been increasingly adopted in dynamic imaging scenarios. These trajectories sample k-space in a spiral pattern, ensuring that all lines pass through the k-space center, thereby enhancing stability in the presence of motion artifacts. Model-based reconstruction, on the other hand, directly estimates $T_1$ maps from k-space data, bypassing the need for intermediate image reconstruction[13]. This approach is particularly appealing as it can integrate motion correction directly into the reconstruction process, reducing the complexity of post-processing. Despite these advantages, radial acquisition presents challenges in maintaining consistent contrast throughout the acquisition, as variations in contrast during sampling can lead to mixed contrast in the final image. Additionally, while model-based reconstruction offers the potential for direct $T_1$ estimation, its application to myocardial imaging requires sophisticated motion-resolved procedures, which can be computationally intensive and complex to implement. Given these considerations, we adopted a strategy that independently reconstructs images and fits $T_1$ maps. This approach balances the need for robust motion handling with the practicality of standard reconstruction workflows, ensuring high-quality $T_1$ maps while maintaining computational efficiency.

For myocardial $T_1$ mapping at 5T, the issue of $B_1+$ inhomogeneity should be seriously considered. At ultra-high fields, interference effects cause severe $B_1+$ inhomogeneity because the wavelength of RF becomes equal to or shorter than the dimensions of typical organs of interest [14], especially in cardiac and abdomen imaging. Therefore, flip angle-related $T_1$ mapping techniques, such as the variable flip angle method [15, 16], may not be suitable. Additionally, due to the high Specific Absorption Rate (SAR) value and susceptibility to $B_0$ inhomogeneity, the conventionally used balanced steady-state free precession (bSSFP) readout is typically replaced by the spoiler gradient echo (GRE) in ultra-high field cardiac imaging.

Considering all the above issues, we rethought the Look-Locker technique that samples the $T_1$ recovery by a continuous GRE acquisition following the IR. The increased $T_1$ at 5T allows a broader acquisition window for sampling the $T_1$ recovery. Furthermore, the Look-Locker technique has low SAR and is less sensitive to the $B_1+$ inhomogeneity due to the inherent low flip angle of GRE, making it suitable for ultra-high fields. Therefore, we propose a multi-inversion recovery Look-Locker method (mIR-rt) for myocardial $T_1$ mapping at 5T. The mIR-rt sequence employs real-time GRE to capture the recovery of inverted magnetization. Diastolic phase images are retrospectively selected based on the time stamps of raw data for $T_1$ fitting. Multi-inversions are used to increase the samples further. Besides, the IR was also optimized for cardiac imaging at 5T to achieve the maximum inversion efficiency. The accuracy of mIR-rt was thoroughly examined through simulations covering a wide range of $T_1$ values. Then the results were validated in a phantom study and 16 healthy volunteers.

## 2. Methods

### 2.1 Sequence

The mIR-rt sequence was implemented on a 5T scanner (Jupiter, United Imaging Healthcare, China). Figure 1 shows its timing diagram. In the sequence, real-time GRE acquisition is performed after an IR pulse to sample the recovery of longitudinal magnetization. After the spins reach the steady state (see Figure S1), a second IR pulse is applied, followed by real-time GRE acquisition to increase the number of data points. Finally, images acquired during diastole are selected for $T_1$ fitting. Specifically, 40 images are acquired between the first and second IR pulses, and 20 images are acquired after the second IR pulse. The first IR is gated by ECG with a trigger delay at diastole. This ensures that initial magnetization recovery is sampled during diastole, which is crucial for fitting accuracy. The total scan time is fixed and does not vary with heartbeats. Parallel imaging is used to reduce the scan time per image so that more data points can be acquired within the limited sampling interval for $T_1$ curve fitting. Since images in systolic are discarded for cardiac $T_1$ fitting, a second IR is applied followed by the real-time GRE to increase the data points.

Temporal Generalized Autocalibrating Partially Parallel Acquisitions (TGRAPPA)[17] is used to reconstruct the k-space data. The calibration k-space lines needed to derive the coil sensitivity maps are obtained by averaging the k-space data of the last ten images. The data acquired at the start of magnetization recovery is not used since the signal intensity varies dramatically. To improve the robustness of $T_1$ estimation, phase-sensitive inversion recovery reconstruction [18] is employed to restore the polarity of MR signals after inversion with the phase of the last image as a reference.

### 2.2 $T_1$ fitting

The $T_1$ fitting model is derived from the Look-Locker method [10]. During real-time GRE

acquisition, the recovery of longitudinal magnetization can be written as a mono-exponential relaxation curve,

$$M(t) = M_\infty - (M_\infty - M(0))\exp(-t/T_1^*), \qquad [1]$$

where $M(0)$ is the initial longitudinal magnetization and $M_\infty$ is the steady state signal of the GRE sequence. $T_1^*$ denotes the apparent relaxation time. Let $M_0$ be the equilibrium magnetization. The inversion recovery of the first IR in mIR-rt sequence starts from $-\delta M_0$, where $\delta$ is the inversion efficiency (InvE). Substituting $M(0) = -\delta M_0$ into Eq.[1], the magnetization after the first IR can be characterized as:

$$M_1(t) = M_\infty - (M_\infty + \delta M_0)\exp(-t/T_1^*), \qquad [2]$$

Similarly, the initial value following the second IR is $-\delta M_\infty$, assuming that the magnetization has reached the steady state of GRE before the second IR. The corresponding magnetization can be written as:

$$M_2(t) = M_\infty - (M_\infty + \delta M_\infty)\exp(-t/T_1^*), \qquad [3]$$

From [10], we know that $m_\infty$ can be approximated by $T_1^*/T_1$ when $T_R \ll T_1^* < T_1$. Substituting $m_\infty$ into Eq.[2] and [3] and combining them, the formula becomes

$$\begin{cases} M_1(t) = M_0^* - (\delta M_0 + M_0^*) \cdot \exp(-t/T_1^*) \\ M_2(t) = M_0^* - M_0^*(\delta + 1) \cdot \exp(-t/T_1^*) \end{cases} \qquad [4]$$

where $M_0^* = M_0 T_1^*/T_1$. Let $A = M_0^* = M_0 T_1^*/T_1$ and $B = M_0(\delta + T_1^*/T_1)$. Eq.[4] can be written as

$$\begin{cases} M_1(t) = A - B \cdot \exp(-t/T_1^*) \\ M_2(t) = A - A(\delta + 1) \cdot \exp(-t/T_1^*) \end{cases} \qquad [5]$$

In the above equation, there are four unknowns A, B, $\delta$, and $T_1^*$ that need to be determined. To improve fitting robustness, the inversion efficiency $\delta$ is estimated by the ratio of image intensities immediately before and after the second inversion pulse first. Then A, B, and $T_1^*$ can be obtained by a three-parameter nonlinear fitting from Eq.[5]. Finally, $T_1$ can be calculated by

$$T_1 = T_1^*(B/A - 1)/\delta, \qquad [6]$$

In real scans, the dummy time $\Delta t$ between the inversion and the start of the acquisition induces a systematic error, and the $T_1$ value should be corrected by [19]:

$$T_{1corrected} = T_1 + 2\Delta t, \qquad [7]$$

From the images acquired during all cardiac phases, diastolic images are selected retrospectively based on timestamps in raw data for $T_1$ calculation. Before curve fitting, the myocardial region of the selected images is aligned using a deep-learning-based image registration method [20]. $T_1$ map is then fitted pixel-by-pixel with Eq.[5].

## 2.3 Simulations

The InvE for a variety of IR designs was calculated using the Bloch simulation with $T_1$ = 1500 ms and $T_2$ = 40 ms. The simulation was conducted with the mri-rf package in the Michigan Image Reconstruction Toolbox (MIRT) [21]. The equilibrium magnetization $M_0$ was set to 1, and InvE was calculated as the ratio of the simulated longitudinal magnetization after the IR pulse to $M_0$. The InvE across the $B_0$ and $B_1$ imperfection ranges were averaged and the pulse parameter combination with the highest average InvE was identified as the optimized set of parameters.

Bloch simulations were also performed to investigate the performance of the proposed mIR-rt method on $T_1$ estimation accuracy, as well as its sensitivity to variations in InvE, $T_1$, and flip angle (FA). The following scenarios were simulated to assess the dependence of $T_1$ accuracy on these three parameters: (1) FA = 7°, $T_2$ = 40 ms, $T_1$ values ranging from 400-2000 ms (100 ms increments), and InvE ranging from 0.5 to 1.0 (0.1 increment); (2) $T_2$ = 40 ms, InvE = 0.85, $T_1$ values ranging from 400-2000 ms (incremented by 100 ms), and FA ranging from 3° to 11° (incremented by 2°); Simulated heart rate was 75 bpm. Other imaging parameters were the same as in-vivo experiments. The MOLLI sequence with scheme 5(3)3 was also simulated to compare its $T_1$ estimation accuracy with the mIR-rt method. The TIs of MOLLI were 150 ms and 300 ms. Spoiled GRE was employed in MOLLI since conventional used bSSFP readout is sensitive to $B_0$ field inhomogeneity and has a high SAR issue at ultra-high field. The normalized errors of $T_1$ estimation of each sequence were calculated using:

$$\text{Normalized } T_1 \text{ error}(\%) = \frac{(T_1^{est} - T_1^{ref})}{T_1^{ref}} \times 100, \quad [8]$$

where $T_1^{est}$ is the $T_1$ value estimation by mIR-rt or MOLLI and $T_1^{ref}$ is the gold standard.

## 2.4 Phantom study

All experiments were carried out at a 5T MR scanner (Jupiter, United Imaging Healthcare, China). Phantom experiments were performed to evaluate the accuracy of $T_1$ measurements using the mIR-rt sequence. The T1MES phantom [22], made of NiCl2-doped agarose gel with different concentrations to mimic different cardiac compartments, was scanned using a local transmit and 24-channel receiver knee coil. We first standardized its $T_1$ value using the IR-SE sequence with 14 TIs logarithmically spaced from 25 to 3000 ms [23]. Imaging parameters were: TR/TE = 20s/15.6ms, FOV = 200×200 mm, in-plane resolution = 1.56×1.56 mm, and slice thickness = 8 mm. To account for incomplete inversion, $T_1$ values of IR-SE were determined by a three-parameter fitting model ($S(TI_i) = A - B \cdot \exp(-TI_i/T_1)$) [24], and served as the reference standard. The total acquisition time was approximately 10.8 hours.

The $T_1$ value of the T1MES phantom was also measured by mIR-rt and MOLLI sequences. mIR-rt acquired a total of 60 images, comprising 40 images after the first IR pulse and 20 images after the second one. The phase encoding lines of each image are 42. Each image took 162.5ms and the total scan time was 10.2 s. The dummy time Δt was 5.8ms. The acquisition scheme of the MOLLI sequence was 5(3)3 using Grappa with R = 2 and separated calibration line number = 24. Other imaging parameters were: flip angle = 7°, bandwidth = 400 Hz/pixel, TR/TE=3.87/1.61 ms. The simulated heart rate was 75 bpm.

**2.5 In-vivo study**

The technique was also validated in 16 healthy volunteers (13 males and 3 females, aged 24 ± 2 years). This study was approved by the Institutional Review Board (IRB) of Shenzhen Institutes of Advanced Technology, Chinese Academy of Sciences with the number of SIAT-IRB-240415-H0887. This study was conducted in accordance with the principles of the Declaration of Helsinki. The 5T MR scanner uses an 8-channel volume transmit loop array for cardiac imaging [25]. For each subject, $B_1$+ shimming of the heart region was performed using an improved magnitude least squares method Eff-MLS[26] to enhance the uniformity of the $B_1$ field. This method enhanced the uniformity of the $B_1$ field by minimizing the spatial variations in the transmit RF field, thereby improving inversion efficiency and $T_1$ mapping accuracy. The Eff-MLS method allows for more precise control of the RF pulses, leading to better correction of $B_1$ inhomogeneity compared to conventional shimming techniques. A 24-channel phased-array abdominal coil was used for signal reception. mIR-rt and MOLLI were used to obtain $T_1$ maps with ECG gating under breath hold. The trigger delay of mIR-rt was set to half of the RR interval to ensure the acquisition starts in the diastolic phase. Other imaging parameters were the same as in the phantom study. Diastole for each heartbeat is determined by an empirical threshold applied to the RR interval calculated from the timestamps in the raw data. An image is classified as diastolic if its k-space center line is acquired between 50% and 95% of the RR interval. Typically, 26-30 images are left for myocardial $T_1$ map fitting after discarding systolic data. 6 volunteers were scanned twice in two separate MR examinations to test the reproducibility of mIR-rt and MOLLI sequences. To account for potential field inhomogeneities that could affect the inversion efficiency and $T_1$ mapping accuracy, $B_0$ and $B_1$ maps were acquired as part of the routine MRI protocol. $B_0$ maps were obtained using a GRE sequence[27]. This method leverages the phase differences between images acquired at multiple TEs to assess $B_0$ inhomogeneity. The sequence parameters included a sufficiently long TR to allow magnetization recovery, multiple TE values to capture signal decay, and a high bandwidth to minimize chemical shift artifacts. The $B_0$ field distribution was calculated by converting the phase differences between images acquired at different TE values. $B_1$ maps were acquired using a Sandwiched presaturation TurboFLASH (satTFL) sequence[28]. This method minimizes the delay between reference ($S_0$) and prepared ($S_1$) images to reduce $T_1$ sensitivity. A nonselective broadband hyperbolic secant (HS8) pulse was used to ensure robustness against $B_0$ inhomogeneities. The $B_1+$ field distribution was calculated using the arccosine of the ratio of $S_1$ to $S_0$ signals, with adjustments for pulse nonlinearity if necessary. This approach enabled rapid 3D $B_1+$ mapping with minimal $T_1$ bias. These maps were used to assess and correct for field inhomogeneities in the cardiac region.

## 3. Analysis

### 3.1 Phantom study

In the phantom study, the region of interest (ROI) of each tube was manually delineated. The spin-echo $T_1$ values were averaged within each ROI to obtain a reference $T_1$ value ($T_1^{ref}$) for each tube. Tubes with $T_1$ values below 500 ms were excluded from the analysis as this study focuses specifically on tissues with longer $T_1$ values. For each $T_1$ mapping sequence, the average $T_1$ ($T_1^{cal}$) and corresponding standard deviation ($T_1^{std}$) were computed over each tube.

The accuracy of the measured T$_1$ value was defined as the absolute difference between $T_1^{ref}$ and $T_1^{cal}$, and the precision for each tube was defined as $T_1^{std}$ [29].

**3.2 In-vivo study**

For the in-vivo study, calculating InvE for each subject is time-consuming and prone to instability. To address this issue, we employed the strategy used in Shao's paper [30] to measure the InvE. Specifically, 5 subjects were selected from the recruited volunteers to establish the average InvE for native myocardial tissue. For each subject, the InvE was calculated by the ratio of average signal intensities within manually drawn ROIs between the 41st and 40th images, which are captured just before and after the second IR. To enhance the robustness, the ROIs for these two images were placed on the ventricular septum and were as similar as possible. The InvE used in the fitting equation (Eq.[5]) was set to the average InvE derived from the 5 volunteers.

To calculate the global T$_1$ values, endocardial and epicardial contours of the left ventricle were manually traced on the T$_1$ maps by scripts developed in MATLAB (R2023a, MathWorks, MA, USA). The average T$_1$ value within the myocardium was calculated as global T$_1$. The consistency of T$_1$ values measured by mIR-rt and MOLLI was assessed using p-values. The differences in myocardial T$_1$ values measured by mIR-rt and MOLLI were assessed using the Wilcoxon signed-rank test, given the small sample size (n = 16) and potential deviations from normality. Although the Shapiro-Wilk test — recommended for small sample normality assessment[31, 32] — indicated no significant deviation from normality in the pairwise differences (p > 0.05 for all myocardial layers), we adopted the non-parametric Wilcoxon test to ensure robustness and reduce sensitivity to outliers[33-35]. A p-value less than 0.05 was considered statistically significant. Bonferroni correction (n = 3) was applied to account for multiple comparisons. All statistical analyses were conducted using Python 3.12 with the SciPy library. The reproducibility of the T$_1$ mapping sequence was evaluated using Intraclass Correlation Coefficients (ICC)[36] for the apex, middle, and base layers. A two-way random effects model (ICC(2,k), absolute agreement definition) was employed to quantify the consistency between two independent scans. This model accounts for both systematic and random variations between measurements, providing a robust estimate of the reliability of averaged values across raters.

## 4. Results

### 4.1 Simulations

Figure 2 shows the simulation results indicating the variation of $T_1$ value measured by mIR-rt on $T_1$, FA, and InvE. Figure 3(a) illustrates the $T_1$ errors in the estimations obtained by mIR-rt under different InvEs and $T_1$ values. The $T_1$ error reduces with the increase of InvE and $T_1$ values. For the $T_1$ values larger than 1000ms, the error is less than 1% even with a low InvE of 0.5. Figure 3 (b) shows the $T_1$ errors varying with different FAs. Similar to Figure 3(a), when the $T_1$ values larger than 1000ms, the normalized $T_1$ errors of all FAs are less than 0.5%. The increase of FA improves the accuracy of measured $T_1$ but reduces the dynamic range and SNR of the $T_1$-weighted images (see Figure S2), which has an adverse effect on the accuracy of estimated $T_1$ in real scenarios. Considering the tradeoff, we chose FA = 7° in this study. Figure 3(c) and 3(d) present the $T_1$ errors of MOLLI sequence under different InvEs, FAs, and $T_1$ values. The $T_1$ error of MOLLI reduces with the increase of InvE and decrease of FA. MOLLI underestimates $T_1$ measurement in case of long $T_1$ values due to the incomplete recovery.

### 4.2 Phantom studies

Figure 3 shows the $T_1$ maps and bar graph of the $T_1$ values obtained by IR-SE, mIR-rt, and MOLLI sequences in the phantom study. The $T_1$ values of mIR-rt are close to the reference values of IR-SE, while MOLLI's $T_1$ values were much lower than the reference. Table 1 summarizes the accuracy and precision of mIR-rt and MOLLI sequences. We can see that mIR-rt has higher accuracy than MOLLI, while the precision of these two sequences is comparable.

### 4.3 In-vivo studies

The average InvE of 5 healthy volunteers was 0.85 for native myocardium[30, 37]. Therefore, InvE was set to be 0.85 in this study.

Figure 4 shows the representative $T_1$ maps of two volunteers using mIR-rt and MOLLI. As we expected, the $T_1$ values obtained using MOLLI are much lower than those from mIR-rt. Please note that different colorbar ranges are employed for MOLLI and mIR-rt to clearly illustrate the maps. In 16 healthy volunteers, the average native myocardial $T_1$ values measured by mIR-rt were 1553 ± 52 ms, 1531 ± 53 ms, and 1526 ± 60 ms at the apex, middle, and base, respectively. In contrast, the corresponding $T_1$ values obtained with MOLLI (1350 ± 48 ms, 1349 ± 47 ms, and 1354 ± 45 ms) were significantly lower across all three myocardial layer. Similarly, the average blood pool $T_1$ value measured by MOLLI (1718 ± 165 ms) was significantly lower than that derived from mIR-rt (1979 ± 97 ms). Table 2 summarizes the myocardial $T_1$ values measured by mIR-rt and MOLLI sequences across the apex, middle, and base layers. The Wilcoxon signed-rank test demonstrated statistically significant differences between the two methods across all layers ($p = 3.05 \times 10^{-5}$), with mIR-rt consistently yielding higher $T_1$ values. These differences remained significant after applying Bonferroni correction.

Notably, the identical p-values observed across the three myocardial layers are a direct result of the rank-based Wilcoxon test producing identical test statistics due to the highly consistent distribution of pairwise differences across all layers. This outcome is methodologically valid, as identical test statistics naturally produce identical p-values in Wilcoxon testing when paired differences are highly consistent, reflecting the uniform and strong effect observed in our dataset.

To assess the reproducibility of mIR-rt, we randomly selected six volunteers from the initial cohort of 16 participants. These volunteers underwent a repeat scan under identical conditions to evaluate the consistency and reliability of the $T_1$ mapping results. This approach ensures an unbiased assessment of the sequence's reproducibility. Table 3 shows the $T_1$ values of 6 volunteers using mIR-rt sequence from two separate scans, which were used for reproducibility analysis. The calculated ICC values were 0.7516 for the apex (indicating good reproducibility), 0.9777 for the middle (excellent reproducibility), and 0.8201 for the base (good reproducibility). These findings confirm that the mIR-rt sequence used in our study offers high reproducibility, particularly in the middle myocardial slices, reinforcing its potential utility in clinical and research applications of myocardial $T_1$ mapping. The relatively lower ICC values observed in the apex and base layers highlight specific technical challenges in these regions. The apex is particularly susceptible to motion artifacts and partial volume effects, while the base is prone to interference from the complex anatomy of the atrioventricular junction. This pattern of inter-layer variability is consistent with previous reports[38-40]. To further improve measurement reproducibility in these regions, future studies may benefit from the integration of 3D whole-heart imaging combined with motion correction algorithms[41, 42].

In the in-vivo experiments, we found inhomogeneity in myocardial $T_1$ maps, especially in the lateral wall. The main reason for this inhomogeneity is the reduced local $B_1$ values, which lead to significantly lower inversion efficiency than intended. Figure 5 shows the representative $T_1$ maps alongside their corresponding $B_0$ and $B_1$ maps. Notably, the lower $B_1$ value in the lateral wall results in a reduced $T_1$ value in this region (indicated by red arrows), even though the redesigned inversion pulse has mitigated some of the adverse effects of $B_1$ inhomogeneity.

## 5. Discussion

This study demonstrated the feasibility of mIR-rt for accurate native myocardial $T_1$ mapping at 5T. The $T_1$ value obtained from the new method is in good agreement with the value obtained using the standard IR-SE sequence in the phantom study. The use of multiple IRs improves the robustness of $T_1$ fitting and an in-vivo $T_1$ map can be obtained by mIR-rt around 10 seconds.

In myocardial $T_1$ mapping at 7T [7], inversion efficiency is measured by acquiring additional images separately at least $5 \times T_1$ after inversion, which greatly prolongs the acquisition time. Another approach, Instantaneous Signal Loss Simulation (InSiL), involves InvE in the $T_1$ fitting model but still needs to acquire an additional proton density-weighted image. In the mIR-rt method, the second IR pulse allows for the calculation of the InvE without the need for additional acquisitions. Since the spins reach a steady state before the second IR pulse, InvE can be calculated using the images obtained just before and after this pulse. As a result, our method eliminates the requirement for extra images.

SAR is essential for cardiac imaging. SAR monitoring at the 5T scanner is based on a human SAR model database. Specifically, this database is based on fifteen human models, which were simulated using Sim4Life software with anatomical images and actual dimensions of the volume transmitting coil[43]. For each subject in a real scan, an appropriate human SAR mode is selected from the database first according to the subject's information, including weight, age, height, and imaging part of the subject. Then the local SAR can be predicted using this human SAR model and the scanning protocol. This SAR monitoring ensures that the local SAR of the sequence does not exceed IEC standards of 60601-2-33.

The underestimation of MOLLI $T_1$ is attributable to three primary factors: (1) It uses the Look-Locker fitting model, which was initially developed for continuous gradient echo readouts; (2) It assumes complete recovery of longitudinal magnetization before the subsequent inversion pulse, which is not true for long $T_1$ tissues or patients with high heart rates; (3) It presumes an ideal inversion pulse, an assumption that is not accurate, especially at 5T. The proposed sequence improves these limitations by (1) Utilizing continuous GRE readouts, which align more closely with the Look-Locker model compared to MOLLI; (2) Assuming that spins reach a steady state of GRE readout before the second IR, a condition that is more feasible than full recovery, especially for long $T_1$ tissues; and (3) Incorporating the effect of inversion efficiency in mIR-rt, which further improves the accuracy of $T_1$ fitting.

The reproducibility of the mIR-rt sequence was generally good across all cardiac regions, with excellent performance in the intermediate layers. However, the accuracy of the vertices was somewhat lower, likely due to the unique technical challenges associated with this area. The apex region is particularly susceptible to motion artifacts and partial volume effects, as it is characterized by a thin wall structure and increased mobility during the cardiac cycle. Additionally, the complex anatomy of the apex and its susceptibility to $B_1$ field inhomogeneity further contribute to reduced accuracy in $T_1$ mapping. While the $B_1+$ shimming technique using the Eff-MLS method has improved overall $B_1$ field uniformity, the apex remains a technically challenging area due to its geometric and physiological characteristics. Future improvements could focus on optimizing $B_1+$ shimming specifically for the apex region or incorporating advanced motion correction algorithms to mitigate motion artifacts in this area. These efforts could further enhance the accuracy and reliability of $T_1$ mapping in the apex, thereby improving the overall diagnostic utility of the technique.

This study has several limitations. First, the precision of myocardial $T_1$ maps at 5T is reduced compared to 3T, as GRE readouts are used instead of bSSFP. Previous studies have also demonstrated this difference in $T_1$ mapping precision between GRE and bSSFP readout[44]. Applying a denoising filter during image processing can improve $T_1$ value precision, a technique often used in commercial sequences. We plan to incorporate it into our post-processing pipeline for future improvements. Second, the temporal resolution of mIR-rt is somewhat limited due to the restricted acceleration rate of TGRAPP, which constrains the number of images available for $T_1$ fitting. To increase the number of fitting images, we utilized a wider diastolic selection window (approximately half of each cardiac cycle), typically selecting 26-30 images for fitting. Spatial variations of the heart region among these images were corrected using the image registration method. Since the acquisition time for the mIR-rt sequence is constant, patients with higher heart rates cover more heartbeats and still have enough images for $T_1$ fitting, despite fewer samples per heartbeat. We have verified that heart rate has little impact on the $T_1$ values estimated by mIR-rt sequence (see Figure S3). In future work, we aim to employ advanced fast imaging techniques, such as deep learning-based methods, to further improve the temporal resolution and image SNR of the sequence.

**6. Conclusion**

The proposed mIR-rt sequence demonstrates superior accuracy compared to MOLLI specifically at 5T. And this study reports myocardial $T_1$ values at 5T for the first time. The optimized IR pulse achieves high inversion efficiency across a broad range of $B_0$ and $B_1$ fields, resulting in enhanced quality of the myocardial $T_1$ map. Furthermore, improving local $B_1$ homogeneity in the heart region at ultra-high fields is crucial, as $B_1$ field inhomogeneity significantly impacts inversion efficiency and, consequently, the uniformity of myocardial $T_1$ values.

**Funding**

This work was supported by the National Natural Science Foundation of China under grant nos. 62322119, 12226008, Shenzhen Science and Technology Program under grant no. RCYX20210609104444089, JCYJ20220818101205012. Also supported by the Key Laboratory for Magnetic Resonance and Multimodality Imaging of Guangdong Province under grant no.2023B1212060052.

imaging for native myocardial T1 mapping using the slice-interleaved T1 mapping (STONE) sequence *NMR in Biomedicine* **29** 1486-96

**Table 1**: Accuracy and Precision of $T_1$ values in four tubes with the two $T_1$ mapping sequences of mIR-rt and MOLLI in phantom study.

|  | Ref (ms) | Accuracy (ms) |  | Precision |  |
| --- | --- | --- | --- | --- | --- |
| No. | IR-SE | mIR-rt | MOLLI | mIR-rt | MOLLI |
| 1 | 1235 | 76 | 166 | 12.45 | 8.47 |
| 2 | 1485 | 120 | 231 | 9.31 | 8.29 |
| 3 | 1885 | 125 | 162 | 20.64 | 17.32 |
| 4 | 950 | 66 | 86 | 8.87 | 7.87 |

**Table 2:** The statistical results of myocardial $T_1$ values using mIR-rt and MOLLI.

| Layer | mIR-rt (Mean ± SD) | MOLLI (Mean ± SD) | Difference (Mean) | p-value (Wilcoxon test) |
|---|---|---|---|---|
| Apex | 1553 ± 54 | 1349 ± 48 | 204.4 | $3.05 \times 10^{-5}$ |
| Middle | 1530 ± 60 | 1350 ± 46 | 179.6 | $3.05 \times 10^{-5}$ |
| Base | 1532 ± 62 | 1352 ± 45 | 180.1 | $3.05 \times 10^{-5}$ |

**Table 3**: The $T_1$ values using mIR-rt sequence from two separate scans for reproducibility analysis.

|  | First Scan (ms) | | | Second Scan (ms) | | |
| --- | --- | --- | --- | --- | --- | --- |
| **Subject No.** | **Apex** | **Middle** | **Base** | **Apex** | **Middle** | **Base** |
| 1 | 1574 | 1557 | 1529 | 1584 | 1570 | 1541 |
| 2 | 1519 | 1497 | 1542 | 1523 | 1506 | 1512 |
| 3 | 1534 | 1501 | 1590 | 1519 | 1510 | 1586 |
| 4 | 1577 | 1554 | 1518 | 1549 | 1558 | 1551 |
| 5 | 1520 | 1531 | 1574 | 1587 | 1528 | 1555 |
| 6 | 1505 | 1522 | 1510 | 1516 | 1523 | 1539 |
| **Average** | 1538 | 1527 | 1544 | 1546 | 1533 | 1547 |

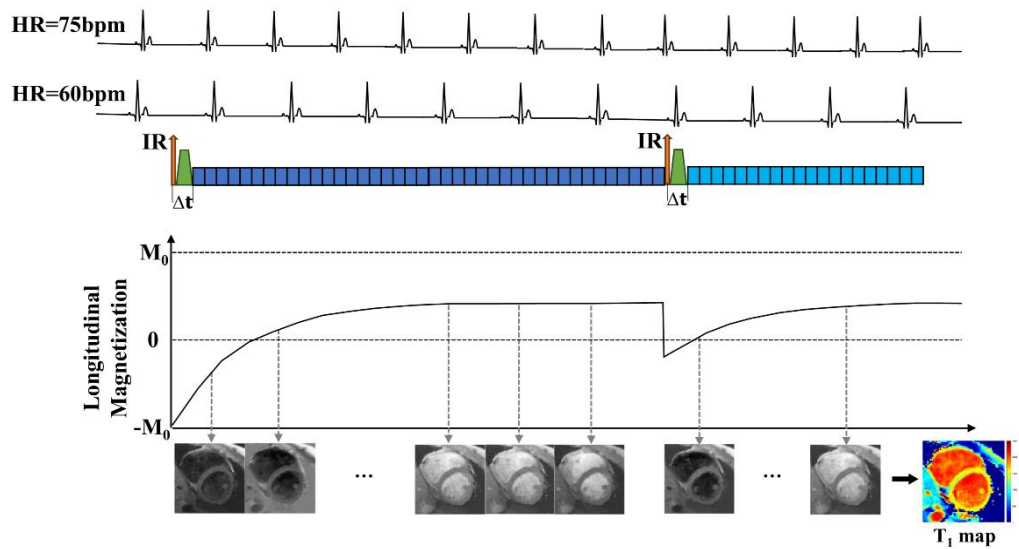

**Figure 1.** The time diagram of the mIR-rt sequence. The sequence has a fixed acquisition time of 10.2 seconds, with the number of cardiac cycles covered depending on heart rate (more cycles for higher heart rates). The top curves show ECG traces for different heart rates, with red arrows indicating inversion pulses (IR). The green trapezoidal blocks represent the spoiler gradient between the IR pulse and the first K-space line, while blue bars indicate image acquisition. Two IR pulses are applied in a single breath-hold, followed by real-time GRE acquisition. Diastolic images are selected and fitted to generate the $T_1$ map using the Look-Locker model. Signal intensity increases as longitudinal magnetization partially recovers from $-M_0$, approaching a pseudo steady-state before the next IR pulse. SNR increases during early recovery but stabilizes after reaching the steady state, ensuring consistent image quality for accurate $T_1$ mapping.

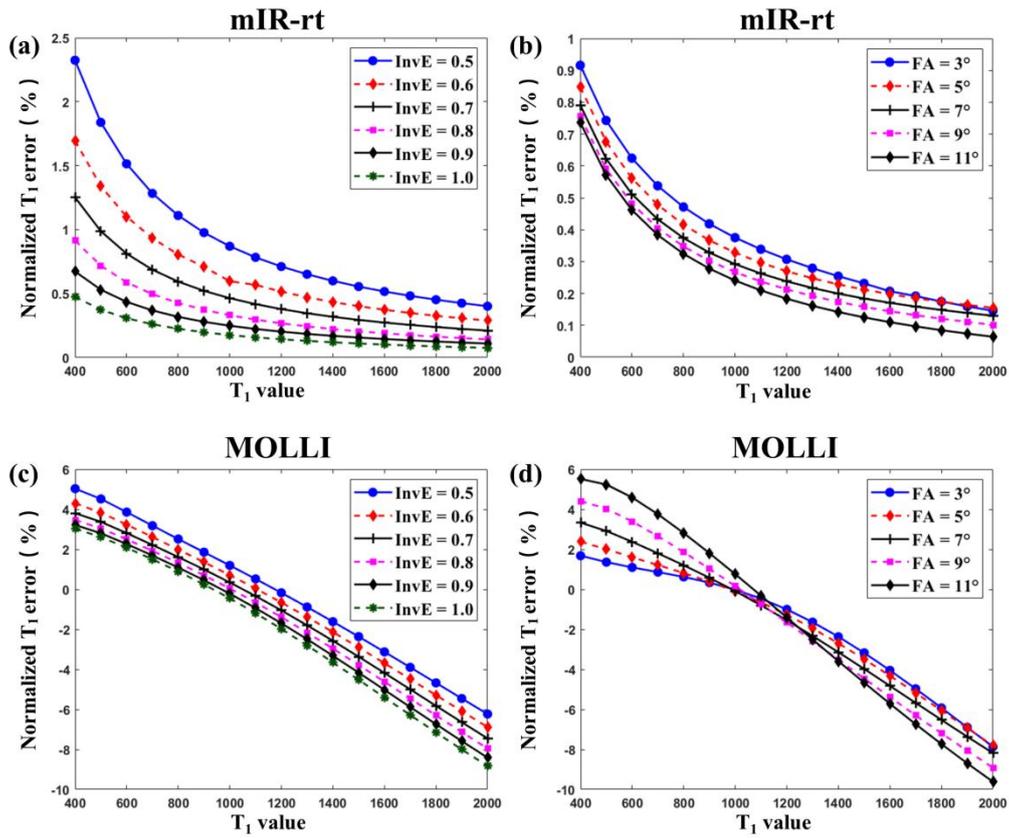

**Figure 2**. Simulation Results (a) $T_1$ estimation errors of mIR-rt under different InvEs and $T_1$ values; (b) $T_1$ errors of mIR-rt varying with different FAs; (c) $T_1$ estimation errors of MOLLI under different InvEs and $T_1$ values; (d) $T_1$ errors of MOLLI varying with different FAs. The normalized $T_1$ errors of mIR-rt are less than 1% for $T_1$ values larger than 1000 ms, while MOLLI underestimates $T_1$ measurement in the case of long $T_1$ due to incomplete recovery.

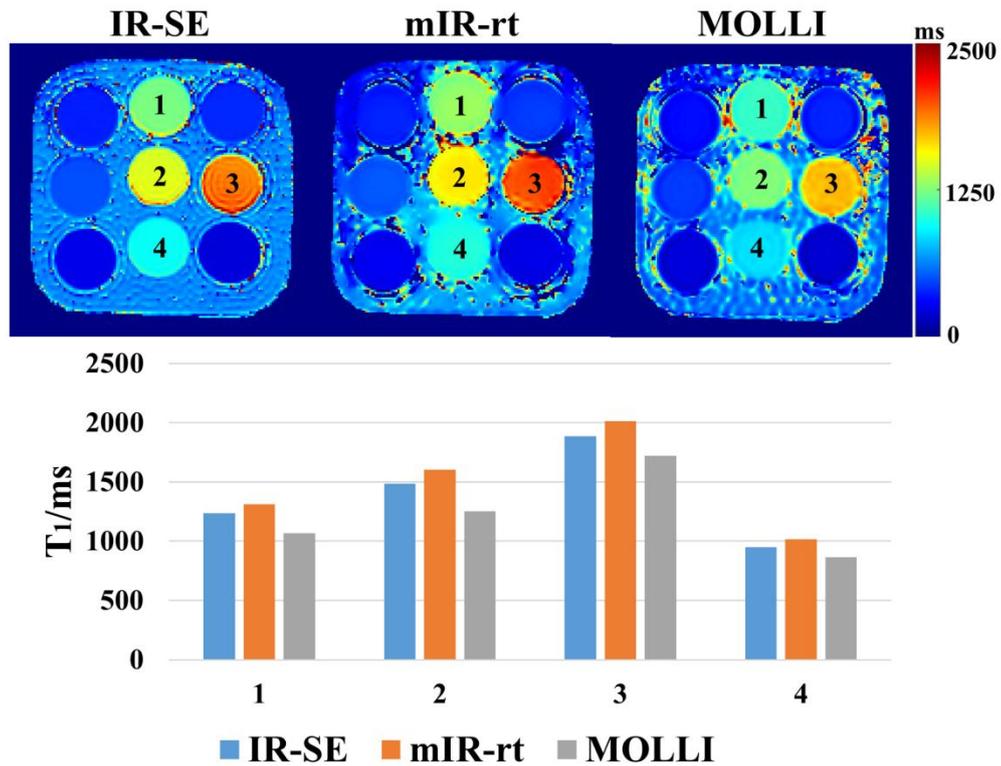

**Figure 3.** The bar graph of $T_1$ values obtained from IR-SE, mIR-rt, and MOLLI sequences in the phantom study. The error lines labeled on the mIR-rt and MOLLI histograms represent the standard deviation. The $T_1$ values of mIR-rt are close to the reference values of IR-SE, while the $T_1$ values of MOLLI are much lower than the reference values, especially for the fifth and sixth tubes with long $T_1$ values.

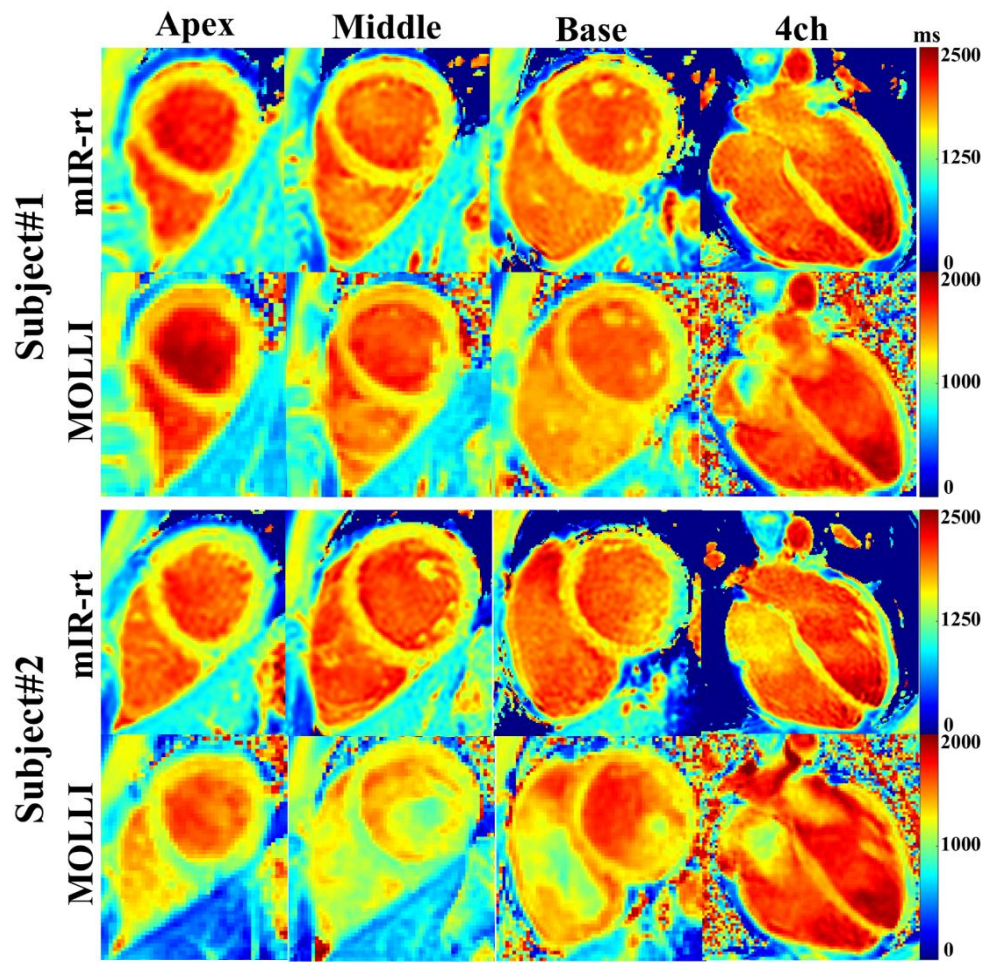

Figure 4. The representative T$_1$ maps of the short-axis and four-chamber from two volunteers using mIR-rt and MOLLI.

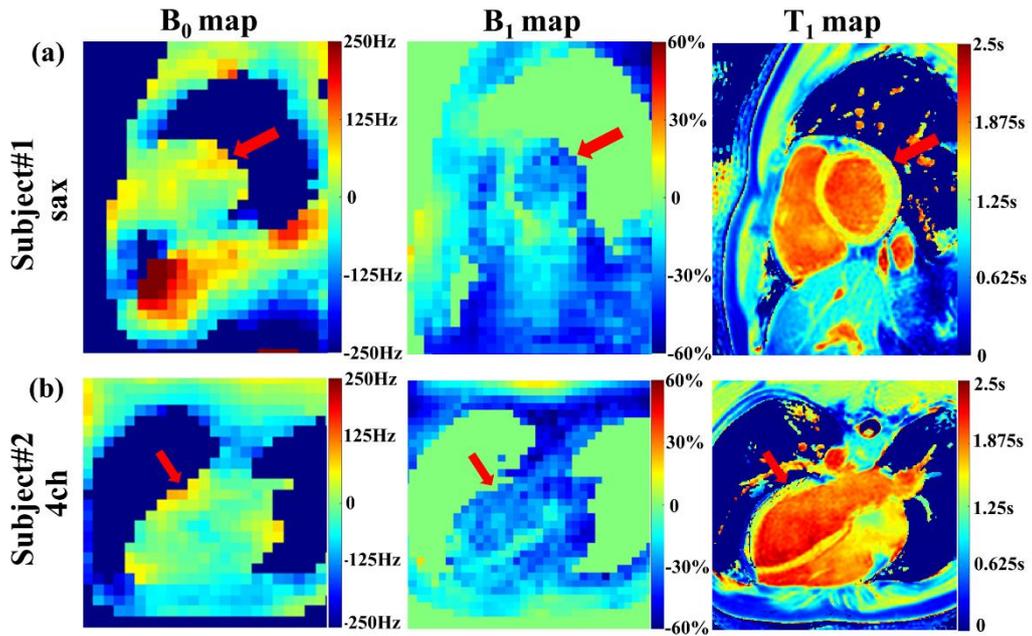

**Figure 5**. The representative $T_1$ maps alongside their corresponding $B_0$ and $B_1$ maps of the short-axis and four-chamber. The red arrows indicate that a lower $B_1$ value on the lateral wall results in a lower $T_1$ value in that area. The $T_1$ maps were calculated using the mIR-rt method.